\documentclass[10pt,letterpaper]{article}
\usepackage[footskip=0.75in]{geometry}

\usepackage{changepage}

\usepackage[utf8]{inputenc}

\usepackage{textcomp,marvosym}

\usepackage{fixltx2e}

\usepackage{amsmath,amssymb}

\usepackage{cite}

\usepackage{nameref,hyperref}

\usepackage{microtype}
\DisableLigatures[f]{encoding = *, family = * }

\usepackage{rotating}

\usepackage[colorinlistoftodos,prependcaption,textsize=tiny]{todonotes}
\reversemarginpar

\raggedright
\usepackage[aboveskip=1pt,labelfont=bf,labelsep=period,justification=raggedright,singlelinecheck=off]{caption}

\makeatletter
\renewcommand{\@biblabel}[1]{\quad#1.}
\makeatother

\date{}

\usepackage{lastpage,fancyhdr,graphicx}
\usepackage{epstopdf}
\fancyheadoffset[L]{2.25in}
\fancyfootoffset[L]{2.25in}
\begin{document}
\vspace*{0.35in}

\begin{flushleft}
{\Large
\textbf\newline{Analysis of Network Clustering Algorithms and Cluster Quality Metrics at Scale}
}
\newline
\\
Scott Emmons\textsuperscript{1,*},
Stephen Kobourov\textsuperscript{2},
Mike Gallant\textsuperscript{1},
Katy B\"orner\textsuperscript{1,3}
\\
\bigskip
\bf{1} School of Informatics and Computing, Indiana University, Bloomington, Indiana, United States of America
\\
\bf{2} Department of Computer Science, University of Arizona, Tucson, Arizona, United States of America
\\
\bf{3} Indiana University Network Science Institute, Indiana University, Bloomington, Indiana, United States of America
\\
\bigskip

%
%





* sremmons@indiana.edu

\end{flushleft}

\section*{Abstract}
Notions of community quality underlie the clustering of networks. While studies surrounding network clustering are increasingly
common, a precise understanding of the realtionship between different cluster quality metrics is unknown. In this paper, we examine the
relationship between stand-alone cluster quality metrics and information recovery metrics 
through a rigorous analysis of four widely-used network clustering algorithms -- Louvain, Infomap, label propagation, and smart local moving.
We consider the stand-alone quality metrics of modularity,
conductance, and coverage, and we consider the information recovery metrics of adjusted Rand score, normalized mutual information,
and a variant of normalized mutual information used in previous work. Our study includes both synthetic graphs and empirical data sets of sizes varying from
1,000 to 1,000,000 nodes.

We find significant differences among the results of the different cluster quality metrics. For example, clustering algorithms can return
a value of 0.4 out of 1 on modularity but score 0 out of 1 on information recovery. We find conductance, though imperfect, to be the stand-alone quality metric that
best indicates performance on the information recovery metrics. Additionally, our study shows that the variant of normalized mutual information used in previous work cannot be assumed to differ only
slightly from traditional normalized mutual information. 

Smart local moving is the overall best performing algorithm in our study, but discrepancies between cluster evaluation metrics prevent us from
declaring it an absolutely superior algorithm. Interestingly, Louvain performed better than Infomap in nearly all the tests in our study, contradicting
the results of previous work in which Infomap was superior to Louvain. We find that although label propagation performs poorly when
clusters are less clearly defined, it scales efficiently and accurately to large graphs with well-defined clusters.


\section*{Introduction}
Clustering is the task of assigning a set of objects to groups (also called classes or categories) so that the objects in the same cluster are more similar (according to a predefined property) to each other than to those in other clusters. 
This is a fundamental problem in many fields, including statistics, data analysis, bioinformatics, and image processing. Some of the classical clustering methods date back to the early 20th century and cover a wide spectrum: connectivity clustering, centroid clustering, density clustering, etc. The result of clustering may be a hierarchy or partition with disjoint or overlapping clusters.
Cluster attributes such as count (number of clusters), average size, minimum size, maximum size, etc., are often of interest.

To evaluate and compare network clustering algorithms, the literature has given much attention to algorithms' performance on ``benchmark graphs''~\cite{girvan_community_2002,lancichinetti_benchmark_2008,lancichinetti_community_2009,ronhovde_local_2010,huang_shrink:_2010}. Benchmark graphs are synthetic graphs into which a known clustering can be embedded by construction. The embedded clustering is treated as a ``gold standard,'' and clustering algorithms are judged on their ability to recover the information in the embedded clustering. In such synthetic graphs there is a clear definition of rank: the best clustering algorithm is the one that recovers the most information, and the worst clustering algorithm is the one that recovers the least information.

However, judging clustering algorithms based solely by their performance on benchmark graph tests assumes that the embedded clustering truly is a ``gold standard'' that captures the entirety of an algorithm's performance. It ignores other properties of clustering, such as modularity, conductance, and coverage, to which the literature has given much attention in order to decide the best clustering algorithm to use in practice for a particular application~\cite{brandes_experiments_2003,schaeffer_survey:_2007,elmqvist_visualizing_2014}.

Furthermore, previous papers that have evaluated clustering algorithms on benchmark graphs have used a single metric, such as normalized mutual information, to measure the amount of ``gold standard'' information recovered by each algorithm~\cite{huang_shrink:_2010,lancichinetti_community_2009,ronhovde_local_2010}. We have seen no studies that evaluate how the choice of information recovery metric affects the results of benchmark graph cluster analysis.

In this paper, we experimentally evaluate the robustness of clustering algorithms by their performance on small (1,000 nodes, 12,400 undirected edges) to large-scale (1M nodes, 13.3M undirected edges) benchmark graphs. We cluster these graphs using a variety of clustering algorithms and simultaneously measure both the information recovery of each clustering and the quality of each clustering with various metrics. Then, we test the performance of the clustering algorithms on real-world network graph data (Flickr related images dataset and DBLP co-authorship network) and compare the results to those obtained for the benchmark graphs. Fig.~1 outlines our entire experimental procedure.


\begin{figure}
\begin{center}
\includegraphics[width=.9\textwidth]{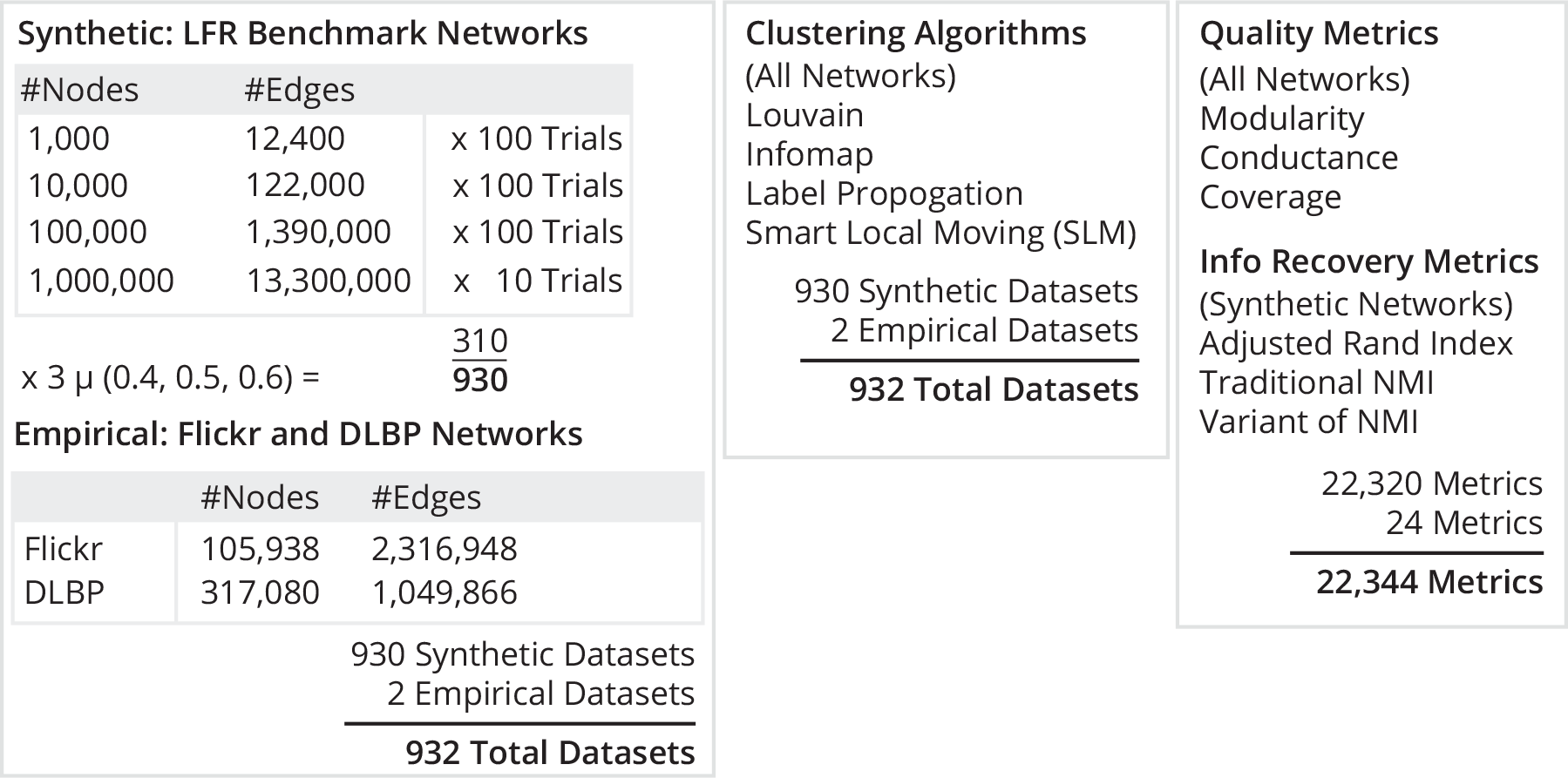}
\label{experimental_procedure}
\end{center}
\caption{The experimental procedure of our clustering algorithm comparison.}
\end{figure}

Specifically, we address the following questions:

\begin{enumerate}
\item How sensitive is a clustering algorithm's performance on benchmark graphs to the choice of information recovery metric?
\item How does a clustering algorithm's performance on the metric of information recovery in benchmark graphs compare to its performance on other metrics such as modularity, conductance, and coverage?
\item How does a clustering algorithm's performance on benchmark graphs scale as the size of the graphs increases?
\item How does an algorithm's performance on benchmark graphs compare to its performance on real-world graphs?
\end{enumerate}

Implementations of all algorithms and all metrics together with links to the synthetic datasets used in this study can be found at \url{http://cns.iu.edu/2016-ClusteringComp} in support of replication and future clustering algorithm comparisons.
\section*{Methods}
\subsection*{Benchmark and Empirical Graphs}
Work on benchmark graphs includes the Girvan-Newman (GN) benchmark~\cite{girvan_community_2002} that consists of 128  nodes grouped into 4 equal-size clusters, for which the internal edges (edges within clusters) exceeds the external edges (edges between clusters). The GN benchmark intuitively captures the idea of a benchmark graph, a graph constructed with an ideal clustering, but it makes no attempt to reflect the structure found in real-world networks which exhibit ``small world'' or ``scale free'' properties~\cite{barabasi_deterministic_2001,leskovec_realistic_2005,watts_collective_1998}.

Lancichinetti et al.~\cite{lancichinetti_benchmark_2008} introduced a new class of benchmark graphs, now known as the the LFR benchmark. This benchmark improves upon the GN benchmark by simulating the properties of networks found in nature with both cluster sizes and node degrees following a power law distribution. These graphs embed a ``gold standard'' clustering by defining a cluster as a set of nodes for which the probability that each node is linked to a node within its cluster is greater than the probability that it is linked to a node outside its cluster, and the LFR graphs have a tunable ``mixing parameter'' $\mu$ that determines the fraction of a node's edges that are external to its assigned cluster. It becomes more difficult to detect clusters as $\mu$ increases, which we illustrate in Fig.~2. Recovering the ``gold-standard'' communities in the LFR benchmark is a greater challenge for clustering algorithms than in the GN benchmark, allowing for more rigorous algorithm testing.


\begin{figure}
\begin{center}
\includegraphics[width=.9\textwidth]{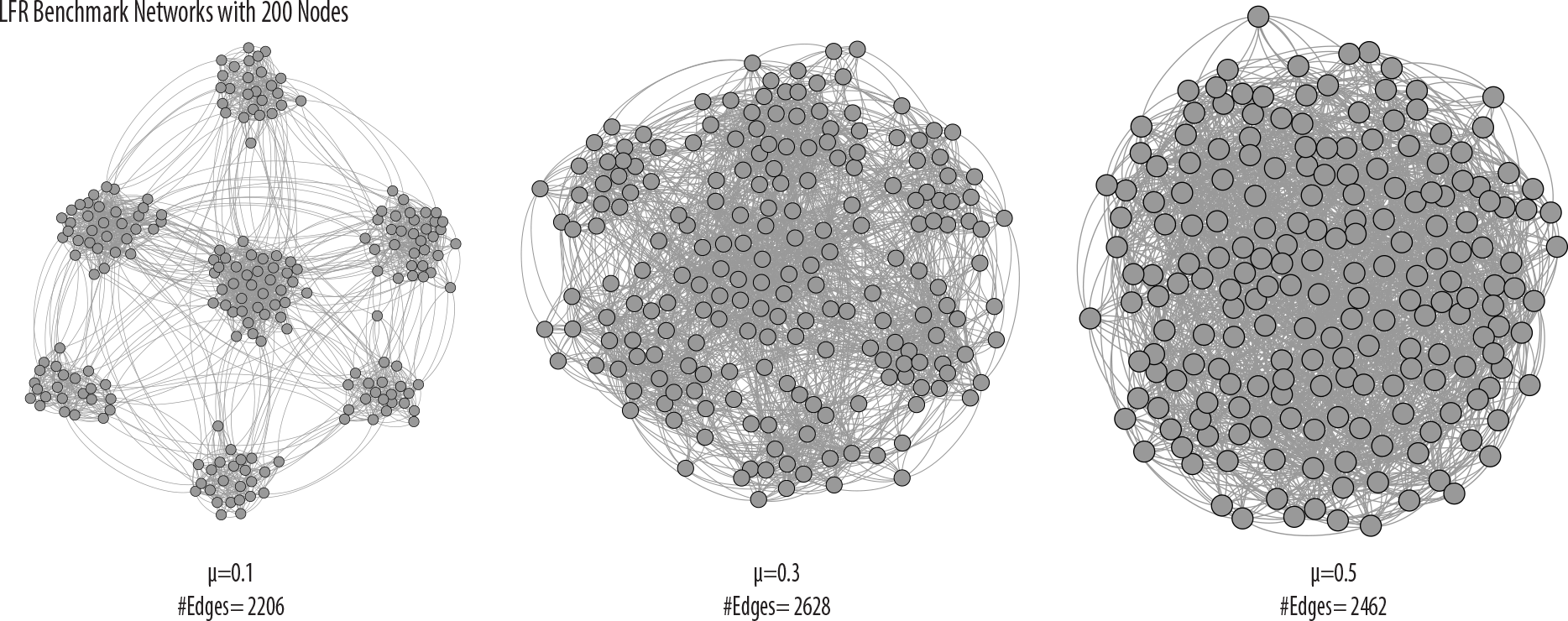}
\label{mu_illustration}
\end{center}
\caption{The impact of $\mu$ on cluster detectability, visualized using the spring-embedded ``ForceAtlas'' algorithm of the software Gephi~\cite{ICWSM09154}.}
\end{figure}

The LFR benchmark has become a standard on which to test algorithms. Lancichinetti and Fortunato used it to compare the performance of twelve clustering algorithms~\cite{lancichinetti_community_2009}, and developers of new clustering algorithms have used it to illustrate competitiveness with existing algorithms~\cite{huang_shrink:_2010,ronhovde_local_2010}. In this study, we use the LFR benchmark for all of our synthetic graphs.

In addition to LFR benchmark synthetic graphs, we also consider real-world graphs to help us gain some intuition about the performance of the clustering algorithms under consideration. Specifically, we use two datasets, one of Flickr related images comprised of 105,938 nodes and 2,316,948 undirected edges~\cite{mcauley_image_2012}, and another of DBLP co-authorships comprised of 317,080 nodes and 1,049,866 undirected edges~\cite{yang_defining_2012}. Both of these datasets are publicly available from the Stanford Network Analysis Project (snap.stanford.edu/data/).
\subsection*{Clustering Algorithms}
Clustering is the task of assigning a set of objects to communities such that objects in the same community
are more similar to each other than to those in other communities. In network clustering, the
literature defines ``similarity'' based on topology. Clustering algorithms seek to capture the intuitive
notion that nodes should be connected to many nodes in the same community (intra-cluster density) but
connected to few nodes in other communities (inter-cluster sparsity). We compare four clustering algorithms
in this study. Each scales to networks of greater than one million nodes.
\paragraph*{Louvain}
The Louvain algorithm~\cite{blondel_fast_2008} is one of the first scalable methods to build on Newman-Girvan
modularity maximization. It is a hierarchical agglomerative method that takes a greedy approach to local
optimization. The algorithm is based on two steps. In the first step, the algorithm iterates over the nodes
in the graph and assigns each node to a community if the assignment will lead to an increase in modularity.
In the second step, the algorithm creates super-nodes out of the clusters found in the first step. The process
repeats iteratively, always using the base-graph to compute the gains in modularity. Although the underlying computational problem is NP-hard, the Louvain algorithm relies on an efficient and effective heuristic that balances solution quality, measured by modularity, and computational complexity,
which, although not precisely known, scales roughly linearly with the number of edges.
\paragraph*{Smart Local Moving (SLM)}
The smart local moving (SLM) algorithm~\cite{waltman_smart_2013} is a more recent modularity optimization
method that has been shown to attain high levels of modularity on graphs with tens of millions of nodes
and hundreds of millions of edges. The algorithm furthers ideas found in the two-step Louvain algorithm and the multilevel refinement method of Rotta and Noack~\cite{rotta_multilevel_2011} by introducing a more advanced local moving heuristic.
For example, the SLM algorithm searches the subgraphs
of identified communities for the opportunity to split the communities for an increase in modularity.
\paragraph*{Infomap}
The Infomap algorithm~\cite{rosvall_maps_2008} is based on the principles of information theory.
Infomap characterizes the problem of finding the optimal clustering of a graph as the problem of finding
a description of minimum information of a random walk on the graph. The algorithm maximizes an objective
function called the Minimum Description Length~\cite{rissanen_paper:_1978,grnwald_advances_2005}, and in practice
an acceptable approximation to the optimal solution can be found quickly. Previous studies have found Infomap's performance to remain stable for networks with up to 100,000 nodes~\cite{lancichinetti_community_2009}.
\paragraph*{Label Propagation}
The label propagation algorithm~\cite{raghavan_near_2007} uses an iterative process to find stable communities in a graph.
The method begins by giving each node in the graph a unique label. Then, the algorithm iteratively simulates a process
in which each node in the graph adopts the label most common amongst its neighbors. The process repeats until the label
of every node in the graph is the same as the label of maximum occurrence amongst its neighbors. While label propagation does not utilize a pre-defined objective function, it is equivalent to a Potts model approach~\cite{tibely_equivalence_2008}.
\subsection*{Cluster Quality Metrics}
A cluster in a network is intuitively defined as a set of densely connected nodes that is
sparsely connected to other clusters in the graph. However, there exists no universal, precise
mathematical definition of a cluster that is accepted in the literature~\cite{lancichinetti_community_2009}.
There are a variety of different metrics that attempt to evaluate the quality of a clustering by capturing
the notion of intra-cluster density and inter-cluster sparsity. Letting $G = (V,E)$ be an undirected graph with adjacency matrix $A$, we use three of the standard cluster quality metrics in our study: modularity, conductance, and coverage. All three are normalized such that scores range from $0$ to $1$, and $1$ is the optimal score.

\paragraph*{Modularity}
The modularity of a graph compares the presence of each intra-cluster edge of the graph with the
probability that that edge would exist in a random graph~\cite{newman_finding_2004,newman_fast_2004}.
Although modularity has been shown to have a resolution limit~\cite{fortunato_resolution_2007}, some of the
most popular clustering algorithms use it as an objective function~\cite{blondel_fast_2008,waltman_smart_2013}.
Modularity is given by Eq.~(\ref{eq:modularity}),
\begin{equation}\label{eq:modularity}
\sum\nolimits_k (e_{kk} - a_k^2)
\end{equation}
where $e_{kk}$, the probability of intra-cluster edges in cluster $S_k$, and $a_k$, the probability of either an intra-cluster edge in cluster
$S_k$ or of an inter-cluster edge incident on cluster $S_k$, are
$$e_{kk} = |\{(i,j) : i \in S_k, j \in S_k, (i,j) \in E\}| / |E|,$$
$$a_{k} = |\{(i,j) : i \in S_k, (i,j) \in E\}| / |E|$$
and where $S_k \subseteq V$. For example, the graph shown in Fig.~3 has modularity equal to
$(\frac{6}{24} - {(\frac{7}{24})}^2) + (\frac{10}{24} - {(\frac{13}{24})}^2) + (\frac{2}{24} - {(\frac{4}{24})}^2) \approx 0.34$.


\begin{figure}
\begin{center}
\includegraphics[width=.6\textwidth]{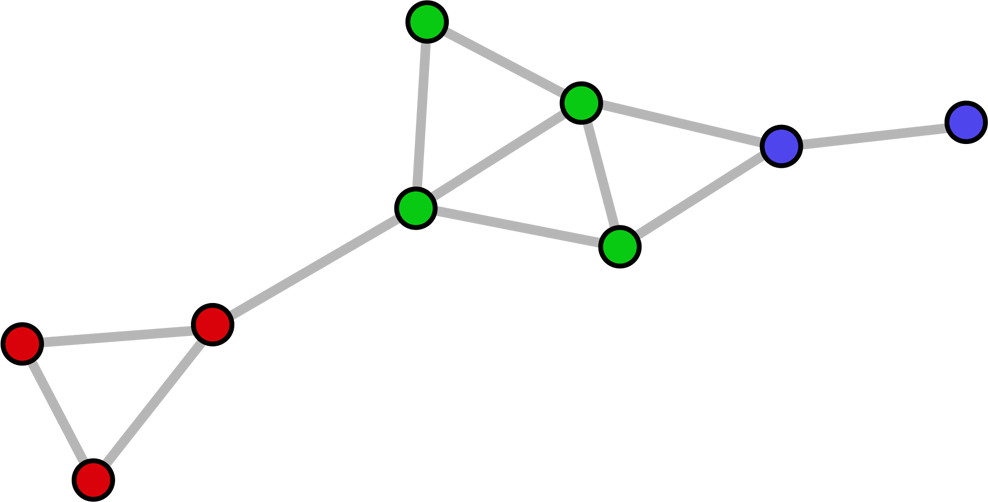}
\end{center}
\label{sample_network}
\caption{A sample network for which modularity $\approx$ 0.34, conductance $\approx$ 0.55, and coverage = 0.75. The color of each node defines its cluster.}
\end{figure}

\paragraph*{Conductance}
We define the conductance of a cluster by the number of inter-cluster edges
for the cluster divided by either the number edges with an endpoint in the cluster or the number of edges that do not have an endpoint in the cluster, whichever is smaller. The conductance for a cluster is given by
Eq.~(\ref{eq:cluster_conductance}),
\begin{equation}\label{eq:cluster_conductance}
\phi (S_k) = \frac{\sum\nolimits_{i \in S_k, j \notin S_k} A_{ij}}{min\{A(S_k),\, A(\overline{S_k})\}}
\end{equation}
where $S_k \subset V$ and $A(S_k) = \sum\nolimits_{i \in S_k} \sum\nolimits_{j \in V} A_{ij} - \sum\nolimits_{i \in S_k} \sum\nolimits_{j \in S_k} A_{ij}$,
the number of edges with an endpoint in $S_k$.

We define the conductance of a graph $G$ to be the average of the conductance for each cluster in the graph, subtracted from $1$.
The conductance for a graph falls in the range $0$ to $1$, and the subtraction makes $1$ the optimal score.
The conductance for a graph is given by Eq.~(\ref{eq:graph_conductance}),
\begin{equation}\label{eq:graph_conductance}
\phi (G) = 1 - \frac{1}{k} \sum\nolimits_k {\phi (S_k)}
\end{equation}

There are several possible ways to define the conductance of a graph that has already been clustered.  In this paper we use inter-cluster conductance as opposed to intra-cluster conductance because the next metric (coverage) deals with intra-cluster density.
Still, it is worth mentioning that this definition of conductance emphasizes the notion of inter-cluster sparsity but does not wholly capture intra-cluster density.
For a more detailed discussion of measures of conductance, including intra-cluster conductance, see Almeida et al.~\cite{almeida_is_2011}.
For example, the graph shown in Fig.~3 would have conductance equal to $1 - \frac{1}{3} (\frac{1}{4} + \frac{3}{7} + \frac{2}{3}) \approx 0.55$.

\paragraph*{Coverage}
Coverage~\cite{elmqvist_visualizing_2014} compares the fraction of intra-cluster edges in the graph to the total number of edges in the graph.
Coverage is given by Eq.~(\ref{eq:coverage}),
\begin{equation}\label{eq:coverage}
\frac{\sum\nolimits_{i,j} A_{ij} \delta (S_i,S_j)}{\sum\nolimits_{i,j} A_{ij}}
\end{equation}
where $S_i$ is the cluster to which node $i$ is assigned and $\delta (a,b)$ is $1$ if $a = b$ and $0$ otherwise.
Coverage falls in the range $0$ to $1$, and $1$ is the optimal score.

While coverage captures the notion of intra-cluster density,
optimizing too heavily for the measure
leads to a trivial clustering in which all nodes are assigned to the same cluster.
For example, the graph shown in Fig.~3 would have coverage equal to $\frac{9}{12} = 0.75$.

\subsection*{Information Recovery Metrics}
When working with an input graph with well-defined clusters, we would like to be able to compare how well a particular clustering algorithm finds the correct clusters. 
It is not trivial to quantify the agreement between the community assignments returned by a clustering algorithm with the ``gold standard''
community assignments embedded in the LFR benchmark graph. Two popular metrics to measure the similarity of clusters are the adjusted Rand
score, which is based on counting, and normalized mutual information, which is based on the Shannon entropy of information
theory~\cite{meila_comparing_2007,vinh_information_2010}. Lancichinetti et al. use a third measure that is a variant
of normalized mutual information for covers~\cite{lancichinetti_community_2009}, or clusterings with overlapping communities. Although it is known that this variant does not agree
with the traditional value of normalized mutual information when no overlap is present, Lancichinetti et al. assume the discrepancy is
negligible~\cite{lancichinetti_community_2009}. We employ all three metrics to study their properties and relationships in benchmark graph analysis,
subsequently describing each in more detail.
\paragraph*{Adjusted Rand Index}
The adjusted Rand index is based on counting. If $X$ and $Y$ are community assignments for each node in the graph, each pair
of nodes $i$ and $j$ can be fit to one of four categories:

$N_{11}$ - $i$ and $j$ are assigned to the same cluster in both $X$ and $Y$

$N_{00}$ - $i$ and $j$ are assigned to different clusters in both $X$ and $Y$

$N_{10}$ - $i$ and $j$ are assigned to the same cluster in $X$ but to different clusters in $Y$

$N_{01}$ - $i$ and $j$ are assigned to different clusters in $X$ but to the same cluster in $Y$

Intuitively, $N_{11}$ and $N_{00}$ indicate agreement between $X$ and $Y$, while $N_{10}$ and $N_{01}$ indicate disagreement between $X$ and $Y$.
The Rand index, often attributed to Rand et al.~\cite{rand_objective_1971}, measures the level of agreement between $X$ and $Y$ as the fraction of agreeing pairs 
of nodes to all possible pairs of nodes, given by Eq.~(\ref{eq:rand_index})~\cite{rand_objective_1971,hubert_comparing_1985},

\begin{equation}\label{eq:rand_index}
RI(X,Y) = \frac{N_{00} + N_{11}}{N_{00} + N_{11} + N_{10} + N_{01}} = \frac{N_{00} + N_{11}}{\binom{N}{2}}
\end{equation}

where $N$ is the number of nodes in the graph.

While the Rand index has a range of $[0, 1]$, chance leads it generally to fall within the more restricted range of $[0.5, 1]$. To correct
for chance, the adjusted Rand index, given by Eq~(\ref{eq:adjusted_rand_index}), was developed~\cite{hubert_comparing_1985,vinh_information_2010}.

\begin{equation}\label{eq:adjusted_rand_index}
\begin{split}
ARI(X,Y) &= \frac{Index - ExpectedIndex}{MaxIndex - ExpectedIndex} \\
         &= \frac{2 (N_{00} N_{11} - N_{01} N_{10})}{(N_{00} + N_{01})(N_{01} + N_{11}) + (N_{00} + N_{10})(N_{10} + N_{11})}
\end{split}
\end{equation}

The adjusted Rand index equals $0$ when the agreement between clusterings equals that which is expected due to chance, and $1$ when the agreement
between clusterings is maximum.
In our experiments we use the Scikit-learn implementation of adjusted Rand score~\cite{pedregosa_scikit-learn:_2011}.
\paragraph*{Normalized Mutual Information}
Normalized mutual information is built on the Shannon entropy of information theory.
Let partitions $X$ and $Y$ define community assignments $\{x_i\}$ and $\{y_i\}$ for each node $i$ in the graph. The Shannon entropy for
$X$ is $H(X) = -\sum\nolimits_x P(x) \log P(x)$, where $P(x)$ is the probability that a node picked at random is assigned to community x.
Likewise, $H(Y) = -\sum\nolimits_y P(y) \log P(y)$ and $H(X,Y) = -\sum\nolimits_x \sum\nolimits_y P(x,y) \log P(x,y)$, where $P(x,y)$ is the probability that a node picked at random is assigned both to $x$ by $X$ and to $y$ by $Y$.
From these entropies of $X$ and $Y$, the mutual information of $X$ and $Y$ is given by $H(X) + H(Y) - H(X,Y)$,
resulting in Eq.~(\ref{eq:mutual_information}).
\begin{equation}\label{eq:mutual_information}
I(X,Y) = \sum\nolimits_x \sum\nolimits_y P(x,y) \log \frac{P(x,y)}{P(x) P(y)}
\end{equation}
The mutual information of $X$ and $Y$ can be thought of as the informational ``overlap'' between $X$ and $Y$, or how much we learn
about $X$ from knowing $Y$ (and about $Y$ from knowing $X$).

In order to normalize the value of mutual information in the range $0$ to $1$, we define the normalized mutual information~\cite{vinh_information_2010} of
$X$ and $Y$ by Eq.~(\ref{eq:normalized_mutual_information}).
\begin{equation}\label{eq:normalized_mutual_information}
I_{norm} (X,Y) = \frac{2 I(X,Y)}{\sqrt{H(X) H(Y)}}
\end{equation}
A normalized mutual information value of $1$ bewteen two clusters denotes perfectly similar clustering, whereas a value of $0$
denotes perfectly dissimilar clustering.
In our experiments we use the Scikit-learn implementation of normalized mutual information~\cite{pedregosa_scikit-learn:_2011}.
\paragraph*{Normalized Mutual Information Variant}
Lancichinetti et al.~\cite{lancichinetti_detecting_2009} created a variant of normalized mutual information to deal with covers,
or clustering assignments in which at least one node is assigned to multiple communities. They let the partitions $X$ and $Y$ define community
assignments $\{x_i\}$ and $\{y_i\}$ for each node $i$ in the graph, where $\{x_i\}$ and $\{y_i\}$ are binary arrays whose lengths
equal the number of different communities in $X$ and $Y$. They denote $x_i^k = 1$ if node $x_i$ is in the $k$th cluster of $X$ and
$x_i^k = 0$ otherwise, ascribing the $k$th entry of $x_i$ to a random variable $X_k$ of probability distribution
$P(X_k = 1) = \frac{n_k}{N}, P(X_k = 0) = 1 - \frac{n_k}{N}$, where $n_k$ is the number of nodes of community $k$ and $N$ is the total
number of nodes in the graph. They denote $y_i$ in the $l$th cluster of $Y$ equivalently.

Lancichinetti et al.~\cite{lancichinetti_detecting_2009} use the probability distributions $X$ and $Y$ to derive the joint probabilities $P(X_k = 1, Y_l = 1)$,
$P(X_k = 0, Y_l = 1)$, $P(X_k = 1, Y_l = 0)$, and $P(X_k = 0, Y_l = 0)$. The additional information of a given $X_k$ to a given $Y_l$,
$$H(X_k|Y_l) = H(X_k, Y_l) - H(Y_l),$$
follows. In calculating the additional information needed to determine $X_k$ from $Y$, Lancichinetti
et al. consider the minimum additional information to determine $X_k$ from all choices of $Y_l$ from $L$ total clusters, yielding
$$H(X_k|Y) = \min\limits_{l\in\{1,2...L\}} H(X_k|Y_l).$$
Dividing by $H(X_k)$ to normalize the expression and averaging the value of each
assignment $k$, from $K$ total clusters, yields Eq.~(\ref{eq:normalized_conditional_entropy}), the normalized entropy of $X$ conditional to $Y$.
\begin{equation}\label{eq:normalized_conditional_entropy}
H(X|Y)_{norm} = \frac{1}{K}\sum\limits_k\frac{H(X_k|Y)}{H(X_k)}
\end{equation}
The symmetric conditional entropy $H(Y|X)_{norm}$ is defined equivalently. Lancichinetti et al.~\cite{lancichinetti_community_2009} use these
conditional entropies to construct their variant of normalized mutual information, given by Eq.~(\ref{eq:normalized_mutual_information_variant}),
subject to an additional constraint discussed in ~\cite{lancichinetti_community_2009} to control for the case of complementary clusterings.
\begin{equation}\label{eq:normalized_mutual_information_variant}
I_{norm} (X,Y) = 1 - \frac{1}{2}[H(X|Y)_{norm} + H(Y|X)_{norm}]
\end{equation}

Note that the normalized mutual information variant of Eq.~(\ref{eq:normalized_mutual_information_variant}), used in the community
detection algorithm comparison of Lancichinetti et al.~\cite{lancichinetti_community_2009}, differs from traditional normalized mutual
information in the case of non-overlapping cluster assignments. It is not clear exactly how different these two metrics are.

\subsection*{Experimental Procedure}
First, we generated a total of 930 undirected LFR benchmark graphs using the parameters outlined in Table~\ref{lfr_parameters}.
We experimented with incrementally increased sizes for the benchmark graphs in powers of ten in order to study the performance of clustering
algorithms as networks scale, and we varied the value of the mixing parameter $\mu$ to study performance as the communities embedded within the graphs become more difficult to recover.
We chose the final LFR benchmark graph generation parameters based on the literature, conversations with other network scientists, and our own experimentation.
We most significantly differ from the parameters of Lancichinetti et al.~\cite{lancichinetti_community_2009} by choosing larger values for the maximum node degree and the maximum community size.
We generated 100 realizations of the LFR benchmark for the three values of the mixing parameter $\mu$ = 0.4, 0.5, and 0.6 for sizes N = 1,000, 10,000, and 100,000, creating 900 graphs.
Additionally, we generated 10 realizations of the LFR benchmark for each of the three values of $\mu$, creating 30 graphs at size N = 1,000,000, the largest size we know of in the literature for a comprehensive algorithm comparison study to date.
While we would have preferred 100 realizations for each value of $\mu$ at N = 1,000,000, the large computational time required to generate the graphs limited us. 

\begin{table}[!ht]
\begin{center}
\caption{
{LFR benchmark graph parameters.}}
\begin{tabular}{|l|l|l|l|}
\hline
\textbf{Param.} & \textbf{Description} & \textbf{Value} & \textbf{Notes} \\ \hline
N & Number of nodes & $[1K\dots 1M]$ & In power of ten increments\\ \hline
k & Average node degree & 25 & Same constant for all sizes\\ \hline
maxk & Maximum node degree & N/10 & To scale with size of graph \\ \hline
$\mu$ & Mixing parameter & {0.4, 0.5, 0.6} & To see impact on performance \\ \hline
$\tau_1$ & Node degree distrib.~exp. & -2 & The default value \\ \hline
$\tau_2$ & Community size distrib.~exp. & -1 & The default value \\ \hline
minc & Min community size & 50 & Same constant for all sizes \\ \hline
maxc & Max community size & N/10 & To scale with size of graph \\
\hline
\end{tabular}
\label{lfr_parameters}
\end{center}
\end{table}

We also used two large real-world graphs obtained from analyzing a dataset of related images on Flickr~\cite{mcauley_image_2012} and a DBLP co-authorship network~\cite{yang_defining_2012}.
These data sets are relatively large in size, but not so large that the task of clustering is computationally infeasible. The Flickr related images network has 105,938 nodes and 2,316,948 undirected edges. The DBLP co-authorships network has 317,080 nodes and 1,049,866 directed edges.
Both of these real-world data sets are available from the Stanford Network Analysis Project (\url{snap.stanford.edu/data/}), enabling reproducibility of our experiments.

Second, we clustered each of the 932 graphs using undirected implementations of the Louvain, smart local moving, Infomap, and label propagation algorithms. We used undirected implementations of all algorithms for consistency with Lancichinetti et al.'s comparison of Louvain and Infomap~\cite{lancichinetti_community_2009}. We also used the lowest hierarchical level of clustering returned by all algorithms, consistent with the study of Lancichinetti et al.\cite{lancichinetti_community_2009} who chose the lowest hierarchical level of Louvain in order to avoid modularity's resolution limit. For each run of Louvain, we used 10 modularity maximization iterations and took the clustering with the greatest modularity. For smart local moving, we used 10 random starts, 10 iterations per random start, and the standard modularity function with a resolution parameter of 1.0. For Infomap and label propagation, we used the default parameters of the implementations at \url{https://sites.google.com/site/andrealancichinetti/clustering_programs.tar.gz}.

We clustered the 930 benchmark graphs using a Dell C6145 cloud server with 4 central processing units, 64 cores, and 256 gigabytes of random-access memory.
We clustered the 2 real-world data sets using Karst, a supercomputer of Indiana University. Karst's compute nodes are IBM NeXtScale nx360 M4 servers. Each contains two Intel Xeon E5-2650 v2 8-core processors, 32 gigabytes of random-access memory, and 250 gigabytes of local disk space.

Finally, we computed the information recovery of the 930 produced clusterings of the benchmark graphs with the embedded gold standard clusterings using the metrics of adjusted Rand index, traditional normalized mutual information, and the variant of normalized mutual information used by Lancichinetti et al. in~\cite{lancichinetti_community_2009}. We calculated the stand-alone quality metrics of modularity, conductance, and coverage for all 932 of the produced clusterings, including both those of benchmark graphs and of real-world graphs.

The code we developed to implement this study, including all scripts, statistics, and analyses, is available and documented at \url{http://cns.iu.edu/2016-ClusteringComp}. 

\section*{Results}


\begin{figure}[!htb]
\begin{center}
\includegraphics[width=.8\textwidth]{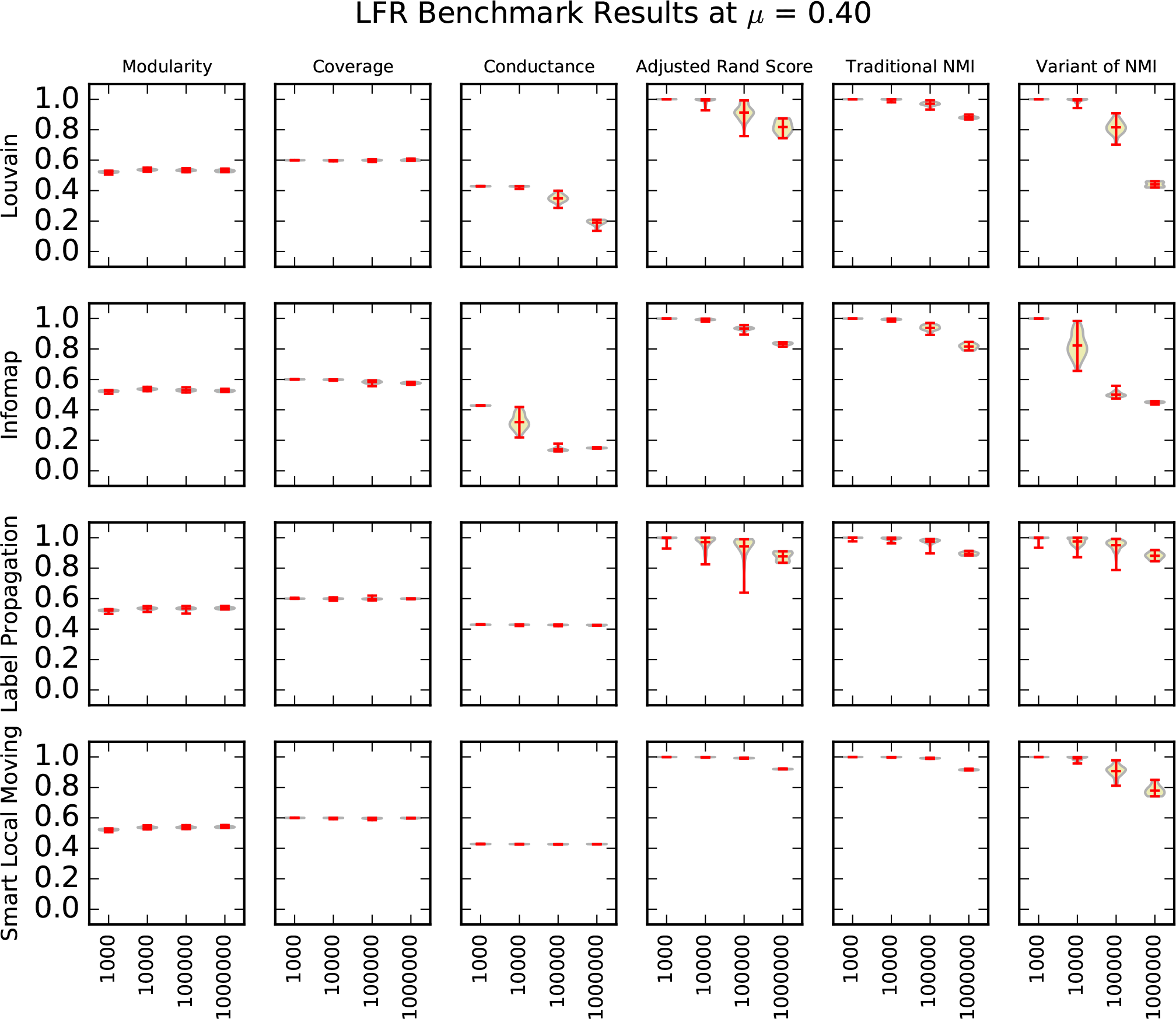}
\end{center}
\label{u40_matrix}
\caption{A matrix of violin plots illustrating the synthetic graph experiment results at $\mu$ = 0.40. We drew each ``violin'' using a Gaussian kernel density estimation. Red lines indicate the minimum, maximum, and mean of the data.
}
\end{figure}


\begin{figure}[!htb]
\begin{center}
\includegraphics[width=.8\textwidth]{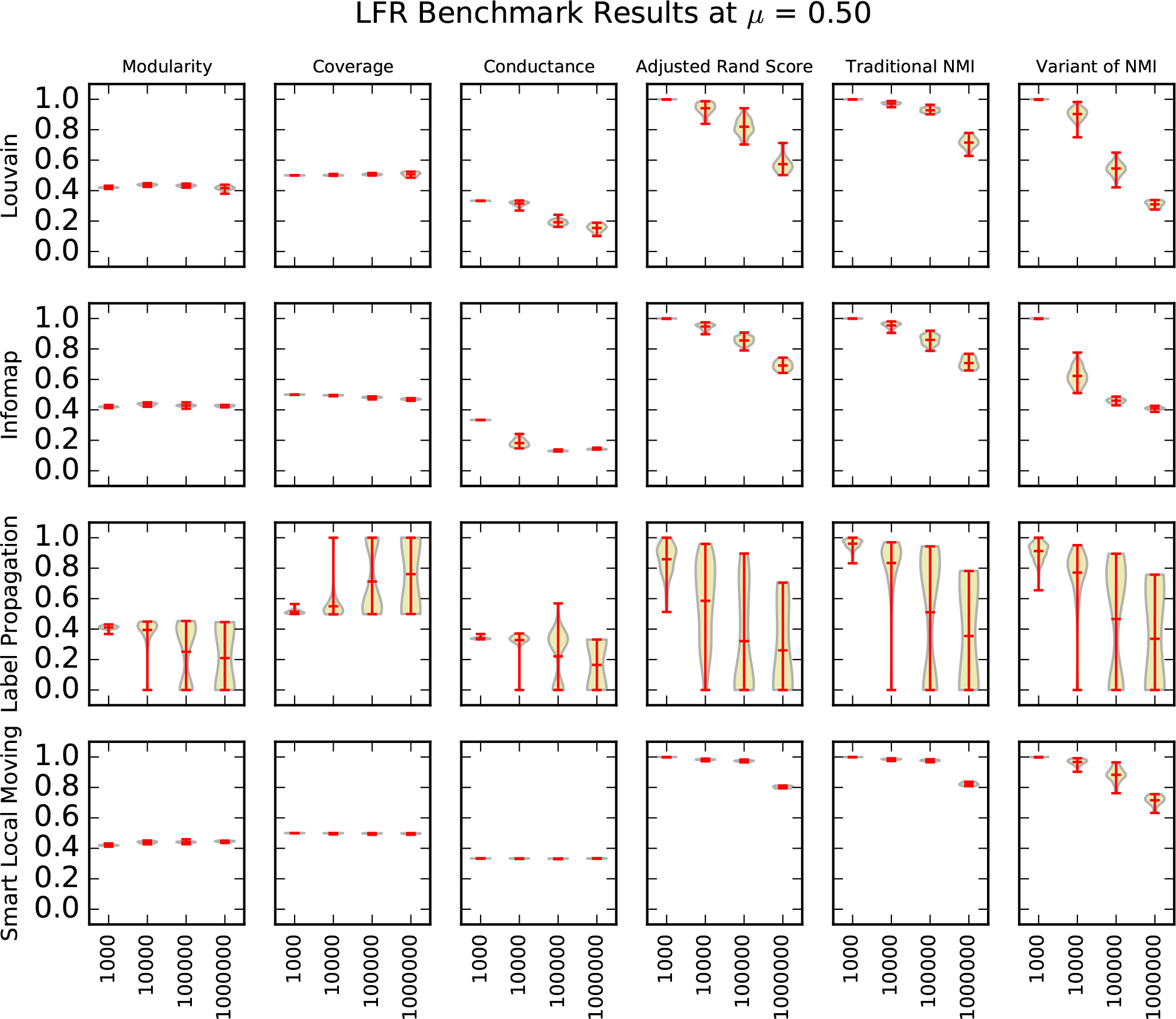}
\end{center}
\label{u50_matrix}
\caption{A matrix of violin plots illustrating the synthetic graph experiment results at $\mu$ = 0.50. We drew each ``violin'' using a Gaussian kernel density estimation. Red lines indicate the minimum, maximum, and mean of the data.}
\end{figure}


\begin{figure}[!htb]
\begin{center}
\includegraphics[width=.8\textwidth]{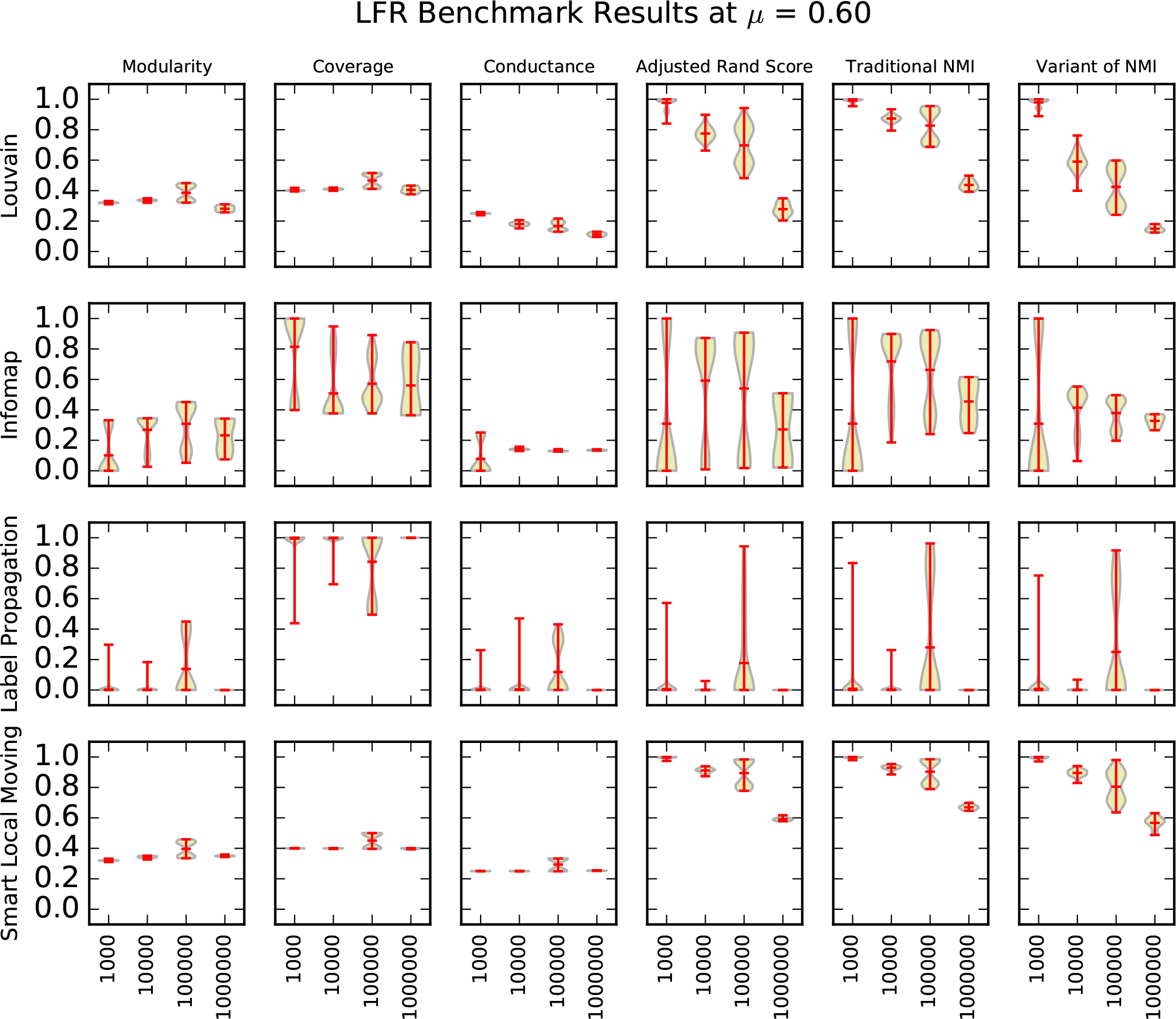}
\end{center}
\label{u60_matrix}
\caption{A matrix of violin plots illustrating the synthetic graph experiment results at $\mu$ = 0.60. We drew each ``violin'' using a Gaussian kernel density estimation. Red lines indicate the minimum, maximum, and mean of the data.
}
\end{figure}

\subsection*{Overview of Resulting Measurements}

We present our results using violin plots. A violin plot is an adaptation of the box plot that enables viewers to make better inferences about the data shown by capturing sample density in addition to summary statistics such as the min, mean and max values~\cite{hintze_violin_1998,correll_error_2014}. Three matrices of violin plots -- one for $\mu$ = 0.40, $\mu$ = 0.50, and $\mu$ = 0.60 in Fig.~4, Fig.~5, and Fig.~6 -- capture the entirety of our synthetic graph results. We rendered the violin plots with lines for the minimum, maximum, and mean of the data, using a Gaussian kernel density estimation for the violin curve~\cite{scott_optimal_1979}.

For a given matrix, each of the clustering algorithms in our study defines a row, and each of the cluster quality metrics in our study defines a column. In this way, each cell in these matrices is a violin plot of the performance of one clustering algorithm by one cluster quality metric. The structure of these matrices allows one to compare the  performance of different algorithms by scanning the columns, and to compare performance of different metrics by scanning the rows.

A fairly clear overall trend is that performance decreases as $\mu$ and the size of the network graphs increases. We expected performance to decrease as $\mu$ increases because higher values of $\mu$ indicate that the embedded clusters are less well-defined. Previous work also conjectured that increases in the size of the network graphs might significantly impact performance\cite{lancichinetti_community_2009}, but our results are the first that we know to verify this claim at this scale. As we discuss in more detail later in the paper, our LFR benchmark graph generation parameters, the ``resolution limit'', and the ``field-of-view limit'' cause decreased performance as the size of the network graph increases~\cite{fortunato_resolution_2007,schaub_markov_2012}.
\subsection*{Comparison of Cluster Quality Metrics}

Our results show that the choice of information retrieval metric has a significant impact on the performance of algorithms.
For example, at $\mu$ = 0.40 and N = 1,000,000, Lancichinetti's variant of normalized mutual information ranks label propagation the highest while
traditional normalized mutual information and adjusted Rand score rank SLM the highest. Louvain outperforms Infomap
on traditional normalized mutual information but loses on adjusted Rand score; see Fig.~4.
These results indicate that a more careful study of cluster quality metrics is needed.

Lancichinetti's variant of normalized mutual information does not match traditional normalized mutual information
when there is no overlap between clusters, which we expected. Unexpectedly, our results show that the variant can differ from the traditional formulation by as much as 0.4; see Louvain's performance at N = 1,000,000 in Fig.~4. This indicates that the results of Lancichinetti et al.~\cite{lancichinetti_community_2009},
which rely solely on this variant of normalized mutual information, cannot be directly applied to the traditional formulation.                                            

Our results suggest that coverage is a poor cluster quality metric. Although we would expect metrics of cluster quality to decrease as $\mu$ and the difficulty of clustering increase, the coverage of label propagation in Fig.~4, Fig.~5, and Fig.~6 increases as $\mu$ increases. Intuitively, as coverage nears a perfect value of 1.0, the clustering of the graph nears the trivial case in which all nodes are assigned to
the same cluster, which suggests that the clustering is too coarse.                                                 

Our results show that modularity is also an unreliable metric to indicate benchmark graph performance. A clustering algorithm's performance
can deteriorate on information recovery metrics without dropping in modularity. Louvain's performance at $\mu$ = 0.50, shown in
Fig.~5, is an example of this. Interestingly, Louvain and SLM -- the two clustering algorithms that optimize modularity --
do not exhibit this behavior any more than does Infomap, which suggests that they are not 
optimizing the measure too heavily. For example, in Fig.~5 at $\mu$ = 0.50 all three of these clustering algorithms show a similar pattern of performance on modularity.

These results question the validity of using metrics such as coverage and modularity to evaluate an algorithm's clustering performance
when a gold standard is not known. Because coverage and modularity do not reflect performance on benchmark graph tests, these two measures
capture fundamentally different properties of clustering than does benchmark graph testing.

Conductance is the metric that best indicates benchmark graph performance in our experiments. The performance of Louvain and Infomap in Fig.~5 at $\mu$ = 0.50
illustrates this point well. While all three information recovery metrics show a steady decline in performance, conductance is the only
stand-alone metric to decline. However, conductance is still an imperfect representation of the information recovery metrics, and there
are other instances in which it fails to reflect a change in information recovery performance.

\subsection*{Comparison of Clustering Algorithms}

A surprising result of our work is Louvain's performance, which surpasses Infomap's in nearly all of our experiments. This contradicts
the previous work of Lancichinetti et al.~\cite{lancichinetti_community_2009} in which Infomap outperformed Louvain.

The ``resolution limit'' of modularity and the ``field-of-view limit'' of both Louvain and Infomap explain how our choice of a relatively large maximum community size leads to this contradictory result. The resolution limit of modularity is the well-known limitation that modularity has in detecting small communities~\cite{fortunato_resolution_2007}. In our experiments, the resolution limit of modularity works in Louvain's favor because our community sizes are relatively large. Analogously, the field-of-view limit marks an upper limit on the size of communities that Louvain and Infomap can detect~\cite{schaub_markov_2012}. Infomap's lack of a resolution limit causes it to suffer acutely from the field-of-view limit and identify smaller clusters than Louvain identifies. In this way, the resolution limit and the field-of-view limit favor Louvain over Infomap in our experiments with large communities.

Note that while our experiments use the bottom hierarchical level of Infomap, which suffers from the field-of-view limit, Schaub et al. have shown how to overcome the field-of-view limit~\cite{schaub_encoding_2012}. Additionally, Kheirkhahzadeh et al. have shown how to overcome the field-of-view limit with the map equation~\cite{kheirkhahzadeh_efficient_2016}.

Lancichinetti et al.~\cite{lancichinetti_community_2009} conclude that the performance of Infomap does not seem to be affected by size. Our
results show that Infomap scales remarkably well to larger sizes but does suffer some performance loss.
For example, at $\mu$ = 0.50 Infomap falls from a 1.0 mean value of traditional normalized mutual information at N = 1,000 to a mean value of
0.70 at N = 1,000,000; see Fig.~5.

Label propagation shows the widest variability in performance of the four clustering algorithms, which is illustrated by the length of its distribution curve in Fig.~5 at $\mu$ = 0.50.
Label propagation's performance is particularly sensitive to $\mu$. When $\mu$ is low, such as at $\mu$ = 0.40,
label propagation scales to size N = 1,000,000 and outperforms other algorithms such as Louvain and Infomap.
However, at higher values of $\mu$, such as $\mu$ = 0.60 in Fig.~6, label propagation's performance rapidly deteriorates.

Label propagation's relative sensitivity to $\mu$, but relative insensitivity to size, suggests a larger
consequence that the best clustering algorithm to be used in practice for networks of large size
depends on how well-defined the clusters are. If the clusters are well-defined, an algorithm that
performs well on lower values of $\mu$, such as label propagation, should be employed. If the clusters
are less well-defined, algorithms such as Louvain and Infomap are superior.

Smart local moving performs best of the algorithms in our study by far. It has an equal to or higher value
than the other algorithms on traditional normalized mutual information and adjusted rand score on virtually all
benchmark graph sizes at all values of $\mu$ in our tests.

\subsection*{Evaluation on Empirical Data Sets with Unknown Gold Standards}

In order to inform the choice of which clustering algorithm to use in practice, we would like to be able to rank the performance
of clustering algorithms on real-world data sets that do not have a ``gold standard'' clustering using stand-alone quality metrics.
However, our earlier results from the synthetic graph analysis reveal that such an absolute ranking of clustering algorithms based on
stand-alone quality metrics does not exist. There is disagreement on the performance of clustering algorithms both
amongst the different stand-alone quality metrics and between the information recovery metrics and the stand-alone quality metrics.

We are not able to make definitive statements about the superiority of clustering algorithms, but it is possible to compute the stand-alone quality metrics, such as those shown in Fig.~7. For example, we see that for the Flickr data set, although smart local moving is nearly the best performer on modularity, it is the worst performer on conductance.
Similarly, smart local moving is the top performer on modularity but the bottom performer on conductance in the DBLP data set. In data sets without knowledge of ground truth, there is not a well-defined way to resolve this disagreement of metrics.


\begin{figure}
\begin{center}
\includegraphics[width=.8\textwidth]{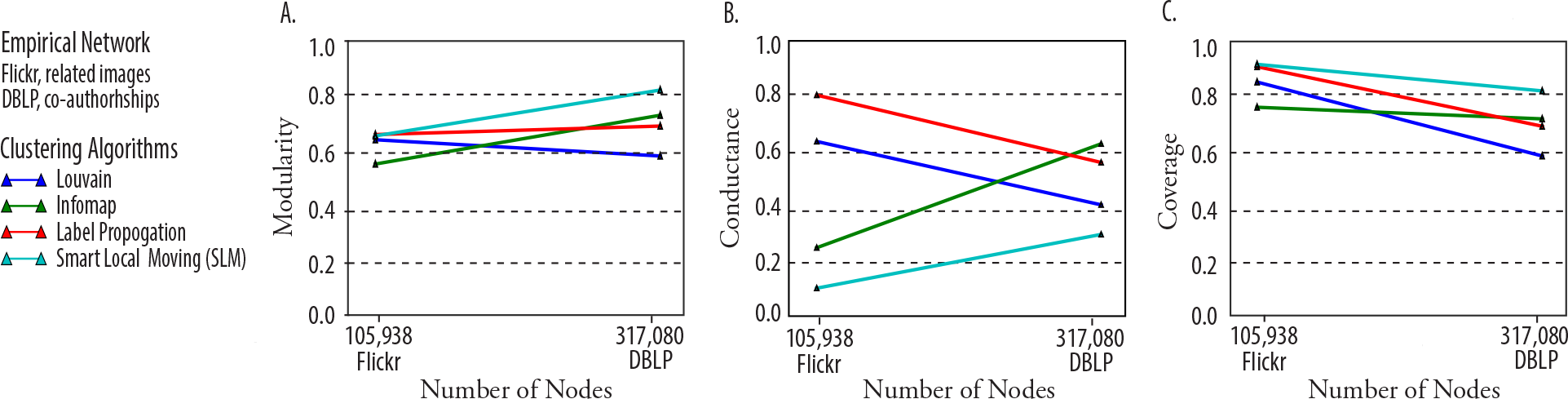}
\end{center}
\label{real_world_graphs}
\caption{(A) A comparison of clustering algorithm performance by modularity on the real-world graphs. (B) A comparison of clustering algorithm performance by conductance on the real-world graphs. (C) A comparison of clustering algorithm performance by coverage on the real-world graphs.
}
\end{figure}

\section*{Discussion}
We evaluate clustering algorithms and cluster quality metrics on graphs ranging from 1,000 to 1M nodes. Our results show overall disagreement
between stand-alone quality metrics and information recovery metrics, with conductance as the best of the stand-alone quality metrics.
Our results show that the variant of normalized mutual information employed by Lancichinetti et al.~\cite{lancichinetti_community_2009}
may significantly differ from traditional normalized mutual information.

Overall, smart local moving is the best performing algorithm in our study.
Note that disagreement between stand-alone quality metrics and information recovery metrics prevents us from claiming that smart local moving is
absolutely superior to the other clustering algorithms. Additionally, the high performance of smart local moving on our LFR benchmark graph tests must be taken with a caveat. The LFR benchmark graphs rely on a synthetic model for their construction with assumptions such as a power law distribution of node degrees. There is inherent circularity in judging a clustering algorithm by its performance on benchmark graphs, and smart local moving's high performance on the LFR benchmark graphs shows that it is based on a model similar to that of the LFR model. However, one may still challenge the LFR model, and potential future work includes analyzing models such as ``CHIMERA'' that enable more precise control of network structure than the LFR benchmark~\cite{franke_chimera:_2016}.

Practitioners seeking to use the best clustering algorithm for a particular application must rely on testing of effectiveness in their respective
domain. Lack of a rigorously defined notion of ``community'', which is intuitively appealing but remains in general to be mathematically defined,
is the root of discrepancies amongst stand-alone quality metrics and information recovery metrics. Without a rigorous notion of a community,
which may vary depending on the domain, absolute statements about the superiority of clustering algorithms cannot be made.

Our results suggest future work in unifying various notions of community, as well as precisely quantifying how current notions of community differ. Additionally, better understanding of the significance of cluster quality metric values (e.g., what does it mean when one clustering algorithm scores 0.1 higher than another in modularity?), will enable more meaningful claims based on these metrics.

\section*{Acknowledgments}
We would like to thank Santo Fortunato for suggestions regarding experimental design, Yong-Yeol Ahn for feedback on the choice of clustering algorithms, Ludo Waltman for editing a completed draft of this work and for helping run the smart local moving code, Martin Rosvall for explaining the relative performance of Louvain and Infomap, and Bahador Saket for discussing early drafts of this work.

Our code draws on Lancichinetti's implementatin of the LFR benchmark (\url{https://sites.google.com/site/andrealancichinetti/files/binary_networks.tar.gz}), Lancichinetti's collection of clutsering methods (\url{https://sites.google.com/site/andrealancichinetti/clustering_programs.tar.gz}), Ludo Waltman's implementation of smart local moving (\url{http://www.ludowaltman.nl/slm/}), Lancichinetti's implementation of the normalized mutual information variant employed in~\cite{lancichinetti_community_2009} and described in~\cite{lancichinetti_detecting_2009} (\url{https://sites.google.com/site/andrealancichinetti/mutual3.tar.gz}), and GMap's implementation of modularity, conductance, and coverage (\url{http://gmap.cs.arizona.edu/}).

This research was partially funded by the National Institutes of Health.
This research was supported in part by Lilly Endowment, Inc., through its support for the Indiana University Pervasive Technology Institute, and in part by the Indiana METACyt Initiative. The Indiana METACyt Initiative at IU is also supported in part by Lilly Endowment, Inc.

%
%
%

\bibliography{sth_arxiv_submission}

\end{document}